\documentclass[twocolumn,pra,aps,superscriptaddress,showpacs]{revtex4}
\usepackage{amsmath}
\usepackage{graphicx}

\begin{document}
\title{Half-knot in the spinor condensates}
\author{Yong-Kai Liu}
\affiliation{Department of Physics, Beijing Normal University, Beijing 100875, China}
\author{Shi-Jie Yang\footnote{Corresponding author: yangshijie@tsinghua.org.cn}}
\affiliation{Department of Physics, Beijing Normal University, Beijing 100875, China}
\affiliation{State Key Laboratory of Theoretical Physics, Institute of Theoretical Physics, Chinese Academy of Sciences, Beijing 100190, China}
\begin{abstract}
We present an exact solution to the stationary coupled nonlinear Gross-Pitaevskii equations which govern the motion of the spinor Bose-Einstein condensates. The solitonic solution is a twisted half-skyrmion in the three-dimension (3D) space. By making a map from the Cartesian coordinates to the toroidal coordinates, we demonstrate it is a linked half-unknot with a fractional topological charge.
\end{abstract}
\maketitle

\section{introduction}
Topological objects appear in a variety of branches of physics from the liquid-crystal colloids\cite{Volovik}, the classical field theory\cite{Faddeev,Battye}, and the optics\cite{Irvine,Dennis,Kivshar}, to the condensed matter physics\cite{Babaev,Yuki2} and the modern universe theory\cite{Vilenkin}. In the dilute atomic gas, quantized vortices and skyrmions have been widely studied and experimentally observed. In particular, the spinor Bose-Einstein condensation (BEC) whose spin direction can change dynamically has provided an ideal pilot to model and explore the rich family of topological excitations\cite{Yuki,Ueda}.

The Hamiltonian of the spinor BEC has the $SO(3)\times U(1)$ symmetry. In contrast to $^4He$ superfluid where only the gauge symmetry is broken, the dilute atomic spinor BEC can also break the spin-rotational symmetry. A large amount of theoretical works has been carried out to study the nontrivial topological structures such as the fractional vortices, the skyrmions, the monopoles and the knots as well\cite{ND,PW,TM,AE,LE,Martikainen,Pietila}. It is known that two kinds of spin textures, the Anderson-Thoulouse skyrmion and Mermin-Ho skyrmion, exist in 2D multi-component BECs. In 3D spinor BECs, 3D skyrmions, as well as knots are theoretically predicted and numerically simulated.

According to the homotopy theory, the topological excitations are classified by the broken symmetry of the corresponding ground states\cite{Yuki,Ueda,MH2}. They are characterized by the topological invariants. In the antiferromagnetic spinor BEC, the order parameter manifold is $M=S^2\times U(1)/Z_2$. Here $U(1)$ denotes the manifold of the superfluid phase $\phi$ and $S^2$ is a two-dimensional sphere. The homotopy groups yield to $\pi_n(M)\cong Z$ ($n=1,2,3$). Singular line defects (vortices) and point defects (monopoles) are determined by the first and the second homotopy classes, $\pi_1(M)$ and $\pi_2(M)$, respectively[13,14]. The monopoles are sometimes called singular Skyrmions\cite{Anglin,Savage,Ruostekoski}. The nontrivial third homotopy class $\pi_3(M)$ implies topological objects of knots. The corresponding topological invariant is called the Hopf charge which counts the number of times the 2D sphere covered by the 3D sphere.

The mean-field order parameter of the spinor BEC is governed by the coupled Gross-Pitaevskii equations (GPE). Attempts for seeking analytical solutions to the coupled GPEs are usually frustrated by two obstacles: one is the nonlinear density-density interactions; the other is the nonlinear spin-spin interactions. In this paper, we present exact solitonic solutions to the stationary GPEs for the $F=1$ BEC in a uniform magnetic field. The solitonic state represents a twisted half-skyrmion in the 3D space and is demonstrated to be a linked half-unknot with a fractional topological charge. To our knowledge, this is the first analytical result to the coupled nonlinear equations in the 3D cartesian coordinates systems (periodical in the $z$-direction). Previous works are mainly in lower-dimensional space or/and by means of numerical simulations, in which the rich possibility of intriguing topological structures are often elusive\cite{Yang1}.

The paper is organized as follows. In Sec.II we describe the analytical method. In Sec.III the topological structure of the stationary state is explored. A summary is included in Sec.IV.

\section{method}
The mean-field Hamiltonian reads,
\begin{eqnarray}
{H}=&&\int d{\bf r} \sum_{m=0,\pm 1}{\psi}_m^* [-\frac{\hbar^2}{2M}\nabla^2-pm+qm^2+V({\bf r})]{\psi}_m \nonumber\\ &&+\frac{1}{2}\int d{\bf r} [c_0\rho^2+c_2{\bf F}^2],
\end{eqnarray}
where the terms associated to $p$ and $q$ represent the linear and the quadratic Zeeman effects, respectively. $\rho(\textbf{r})=\sum_m |\psi_m(\bf r)|^2$ is the total density of the condensate. The spin-polarization vector $\textbf{F}=\sum_m\psi_m^*\hat{\textbf{f}}_{mn}\psi_n/\rho$, with $\hat{\textbf{f}}$ the spin matrices. The nonlinear coupling constants $c_0=(g_0+2g_2)/3$ and $c_2=(g_2-g_0)/3$, where $g_F$ relating to the $s$-wave scattering length of the total spin-$F$ channel as
$g_F=4\pi\hbar^2a_F/M$. By choosing the units of $\hbar=M=1$, the stationary GPEs in a uniform external field ($V({\bf r})=0$) are written as
\begin{widetext}
\begin{eqnarray}
\mu\psi_1=[-\frac{1}{2}\bigtriangledown^2-p+q+(c_0+c_2)(\vert\psi_1\vert^2+\vert\psi_0\vert^2)+(c_0-c_2)
\vert\psi_{-1}\vert^2]\psi_1+c_2\psi_0^2\psi_{-1}^* ,\nonumber\\
\mu\psi_0=[-\frac{1}{2}\bigtriangledown^2+(c_0+c_2)(\vert\psi_1\vert^2+\vert\psi_{-1}\vert^2)+c_0
\vert\psi_0\vert^2]\psi_0+2c_2\psi_0^*\psi_1\psi_{-1}, \nonumber\\
\mu\psi_{-1}=[-\frac{1}{2}\bigtriangledown^2+p+q+(c_0+c_2)(\vert\psi_-1\vert^2+\vert\psi_0\vert^2)+(c_0-c_2)
\vert\psi_{-1}\vert^2]\psi_{-1}+c_2\psi_0^2\psi_1^* .\label{stationary}
\end{eqnarray}
\end{widetext}

We try to decouple the nonlinear density-density interactions by separating the variables and obtain the independent equations in the $x$- and $y$-direction, respectively. The complexity of the spin-spin couplings are surpassed by demanding that the hyperfine states satisfy $\psi_{-1}=-\psi_1^*$ and $\psi_0^*=\psi_0$, which yield the average of the spin-polarization $|\textbf{F}|\equiv 0$. This vanishing spin-polarization state is most easily realizable in the antiferromagnetic BECs. The wavefunction of each hyperfine state is decomposed as $\psi_1=[AX_1(x)Y(y)+iBX(x)Y_1(y)]Z(z)$ and $\psi_0=DX(x)Y(y)$, where $X(x)$, $X_1(x)$, $Y(y)$, and $Y_1(y)$ are real functions and $Z(z)$ is a complex function. $A$, $B$, and $D$ are real constants with $D^2=2(A^2+B^2)$. If we seek the solution that the density is uniform in the $z$-direction, i.e., $|Z(z)|^2=1$, then the total density is written as $\rho({\bf r})=2(B^2X^2(x)+A^2Y^2(y))$. To this purpose, it requires that $X_1^2(x)+X^2(x)=1$ and $Y_1^2(y)+Y^2(y)=1$, which are fulfilled by the hyperbolic functions or the Jacobian elliptical functions\cite{Cong}.

The 2D problem is then converted into a set of 1D differential equations, which can be solved self-consistently by making use of the properties of hyperbolic functions. The 3D solution will satisfy the periodic boundary condition in the $z$-direction.

\section{half-knot soliton}
The 3D solitonic form of solutions is constructed as
\begin{equation}
\Psi=\left(
          \begin{array}{c}
            -\frac{1}{\sqrt{2}}[\textrm{tanh}(kx)\textrm{sech}(ky)-i\textrm{sech}(kx)\textrm{tanh}(ky)]e^{-ik_zz} \\
           \sqrt{2}\textrm{sech}(kx)\textrm{sech}(ky) \\
            \frac{1}{\sqrt{2}}[\textrm{tanh}(kx)\textrm{sech}(ky)+i\textrm{sech}(kx)\textrm{tanh}(ky)]e^{ik_zz} \\
          \end{array}
     \right),\label{solution}
\end{equation}
where the periodic boundary condition in the $z$-direction requires $k_z=2n\pi/L_z$, with $n$ the number of modes. For simplicity, we have chosen $k_x=k_y\equiv k$. The total density of the condensate $\rho(\textbf{r})=\textrm{sech}^2(kx)+\textrm{sech}^2(ky)$ is uniform in the $z$-direction, which endues the Eqs.(\ref{stationary}) be decoupled and the variables be separable according to the dimensions. This solution has solitonic form in the $x-y$ plane. The parameter relations are straightforwardly obtained by substituting the state (\ref{solution}) into the Eqs.(\ref{stationary}) as $c_0=-k^2$, $p=0$, $q=-\frac{1}{2}k^2-\frac{1}{2}k_z^2$, and $\mu= -k^2$. We note that the quadratic Zeeman energy plays the role of balancing the chemical difference between the hyperfine states.

The 3D order parameter can be rewritten as $\Psi(\textbf{r})=\sqrt{\rho(\textbf{r})}\chi(\textbf{r})$, where the $\chi(\textbf{r})$ is a normalized spinor, $\chi^\dagger(\textbf{r})\chi(\textbf{r})=1$. Instead of the vanishing spin-polarization, we consider a real unit vector ${\bf d}=(d_x,d_y,d_z)^\textrm{T}$, by which the spinor $\chi({\bf r})$ is represented as\cite{Ja,HZ}
\begin{equation}
\chi({\bf r})=\left(\begin{array}{c}
\frac{-d_x+id_y}{\sqrt{2}}\\
d_z\\
\frac{d_x+id_y}{\sqrt{2}}
\end{array}\right).
\end{equation}
The ${\bf d}({\bf r})$ vector is then given by
\begin{widetext}
\begin{equation}
{\bf d}({\bf r})=\frac{1}{\sqrt{\rho({\bf r})}}\left(\begin{array}{c}
\textrm{tanh}(kx)\textrm{sech}(ky)\cos(k_z z)-\textrm{sech}(kx)\textrm{tanh}(ky)\sin(k_z z)\\
\textrm{tanh}(kx)\textrm{sech}(ky)\sin(k_z z)+\textrm{sech}(kx)\textrm{tanh}(ky)\cos(k_z z)\\
\sqrt{2}\textrm{sech}(kx)\textrm{sech}(ky)
\end{array}\right).
\end{equation}
\end{widetext}
In quantum magnetism, $\bf d$ is also called the N\'{e}el vector which corresponds to the staggered magnetization of the antiferromagnetic state.
\begin{figure}[h]
\begin{center}
\includegraphics*[width=8cm]{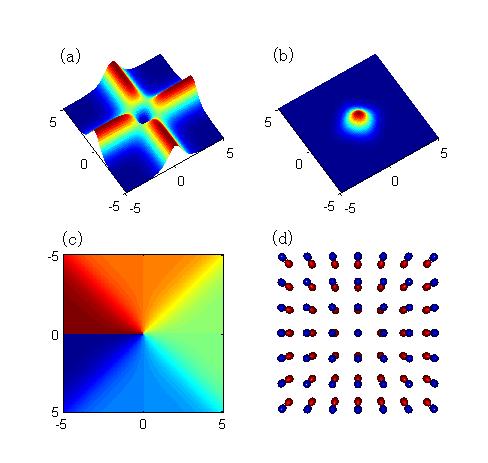}
\end{center}
\caption{(Color online) The 2D density distribution of the state (\ref{solution}) for (a) $\psi_1$ (or $\psi_{-1}$) and (b) $\psi_{0}$. (c) The phase distribution for $\psi_{-1}$. (d) The spherical harmonics representation of the spinor density.}
\end{figure}

\subsection{2D geometry}
To reveal the topological structure of this state, we first examine the 2D situation which is fulfilled by simply setting $z=0$ in the state (\ref{solution}).  Figure 1(a) and (b) display the density distributions of the hyperfine states $\psi_1$ (or $\psi_{-1}$) and $\psi_0$, respectively. It shows that the hyperfine state $\psi_1$ ($\psi_{-1}$) forms a line-shape soliton, whereas $\psi_0$ forms a 2D point-shape soliton. The phase distribution of the $\psi_1$ ($\psi_{-1}$) in Fig.1(c) reveals a vortex (anti-vortex) at the center, where the $\psi_0$ component resides in. Therefore, the spinor BEC forms a coreless solitonic vortex. Fig.1(d) displays the spherical harmonic representation of the spinor condensate density which is expressed by $|\Psi({\bf r},\theta_\textrm{s},\phi_\textrm{s})|^2=|\sum_{m=0,\pm1}\psi_m({\bf r})Y_{1m}(\theta_\textrm{s},\phi_\textrm{s})|^2$, where $(\theta_\textrm{s},\phi_\textrm{s})$ are the coordinates in the spin-space\cite{Ueda}. It exhibits a semi-hedgehog structure. We demonstrate this structure is a half-skyrmion as follows.
\begin{figure}[h]
\begin{center}
\includegraphics*[width=8cm]{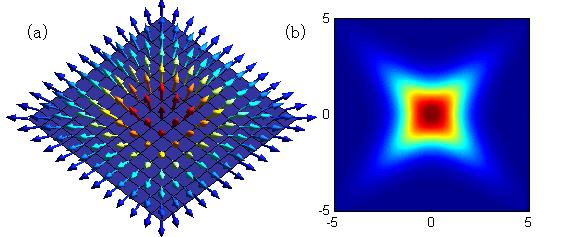}
\end{center}
\caption{(Color online) (a) The 2D texture of the $\textbf{d}$ field. (b) The topological charge density $q(\bf r)$. The total charge is $Q_{2D}=1/2$.}
\end{figure}

In 2D geometry, the vector ${\bf d}({\bf r})=\frac{1}{\sqrt{\rho({\bf r})}}(\textrm{tanh}(kx)\textrm{sech}(ky),\textrm{sech}(kx)\textrm{tanh}(ky),\sqrt{2}\textrm{sech}(kx)\textrm{sech}(ky))^\textrm{T}$. Since $|{\bf d}|=1$ and ${\bf d}_z>0$ everywhere, it forms the surface of a semi-sphere in the order parameter manifold which is denoted as $S^2_{1/2}$ (see Fig.10 in \cite{MM}). Here the subscript $1/2$ indicates {\it half} of a sphere. The vectorial distribution of the ${\bf d}$ field is shown in Fig.2(a). The vector points upward at the center and gradually twists in the 2D plane at large distance. This texture represents a half-skyrmion which covers exactly half a sphere\cite{A}. It is also called a Mermin-Ho Skyrmion or meron which has been studied in the condensed matter physics such as the quantum Hall effects\cite{LB} and high temperature superconductivity\cite{Ta}. It was created experimentally in the superfluid $^3He$-$A$ in a rotating cylinder\cite{VM,R} and in dilute atomic BEC by adiabatic deformation of the magnetic trap\cite{AE}. Fig.2(b) shows the topological charge density $q({\bf r})=\frac{1}{4\pi}\textbf{d}\cdot(\partial_x\textbf{d}\times\partial_y\textbf{d})$, which characterizes the geometric distribution of the spin texture\cite{Ja}. The total charge  $Q_{2D}=\int q({\bf r}) d{\bf r}=+1/2$ is a fractional number\cite{Yang1,WM1}. This is the reason why it is called a half-skyrmion. The anti-half-skyrmion solution with $Q_{2D}=-1/2$ can be obtained by simply taking the complex conjugacy of the state (\ref{solution}).

\begin{figure}[h]
\begin{center}
\includegraphics*[width=8cm]{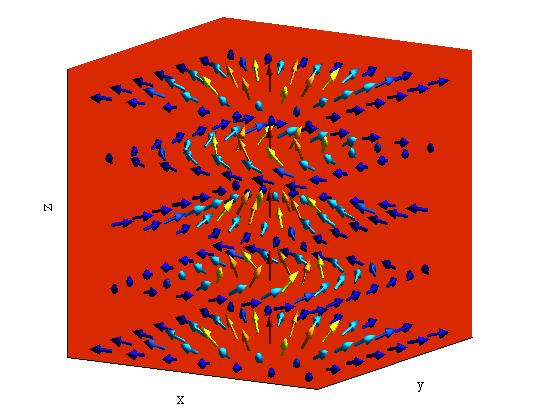}
\end{center}
\caption{(Color online) The 3D texture of the $\textbf{d}$ field.}
\end{figure}

It is notable that the half-skyrmion in the uniform system has an infinite energy, as discussed by L. Brey et al in the quantum Hall system\cite{LB}. This problem is circumvented by considering a pair of half-skyrmion and anti-half-skyrmion in a finite space which has a finite energy. It can be implemented in the periodical solution by simply substituting the hyperbolic functions $\tanh$ ($\textrm{sech}$) with the Jacobi elliptical functions $\textrm{sn}$ ($\textrm{cn}$) and taking the solitonic limit of unit modulus\cite{Cong}. Each cell consists of a half-skyrmion and anti-half-skyrmion pair that is localized in a finite space and consequently has finite energy.

\subsection{3D geometry}
Now we further explore the 3D topological structure of the state (\ref{solution}). The periodic boundary condition in the $z$-direction implies that it essentially forms a twisted vortex ring. Figure 3 display the vectorial distribution of the ${\bf d}({\bf r})$ vector in the 3D space. The vector points from upward at $z=0$ to inplane at large distance while winds around the $z$-direction. The state that has a vanishing average spin-polarization implies it is locally degenerate [14]. The order parameter is invariant under an arbitrary rotation about ${\bf d}$, \textit{i.e.}, $\exp[-i{\bf d}\cdot{\hat{\bf f}}\phi]\Psi=\Psi$, where $\phi$ is the rotation angle. The manifold of the unit ${\bf d}$ vector in the 3D geometry is similar to that in the 2D geometry.
\begin{figure}[h]
\begin{center}
\includegraphics*[width=8cm]{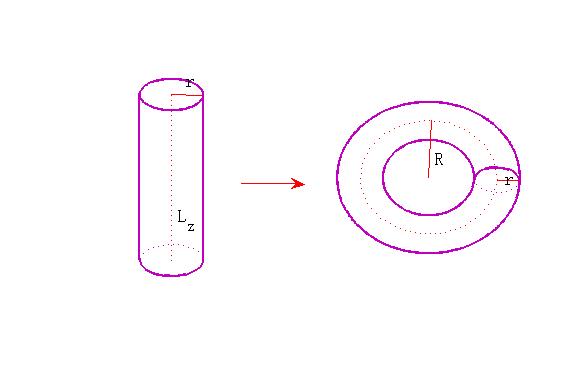}
\end{center}
\caption{(Color online) Schematic description of the mapping (\ref{mapping}) from a cylinder to a torus. The symmetric axis of the cylinder is mapped as the core of the torus.}
\end{figure}

In order to get a more intuitive view of the topological structure, we make a mapping from the Cartesian coordinates to the toroidal coordinates by
\begin{equation}
\left(\begin{array}{c}
x\\
y\\
z
\end{array}\right)\rightarrow
\left(\begin{array}{c}
\tilde x\\
\tilde y\\
\tilde z
\end{array}\right)=\left(\begin{array}{c}
(x+R)\cos(z/R)\\
(x+R)\sin(z/R)\\
y
\end{array}\right),\label{mapping}
\end{equation}
where $R=L_z/2\pi$ defines the size of the torus. This mapping transforms a cylinder into a torus by reconnecting its two ends, as schematically shown in Fig.4. $|{\bf\tilde{r}}|=R$ indicates the position of the toroidal core. In terms of vector ${\bf d}$ it twists $n$-times around the core before joining the ends, implying it is a twisted vortex ring.
\begin{figure}[h]
\begin{center}
\includegraphics*[width=8cm]{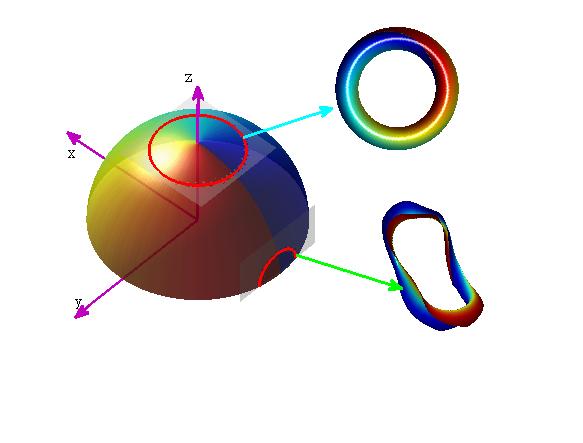}
\end{center}
\caption{(Color online) A loop (string) on the $S^2_{1/2}$ corresponds to a twisted tube (Mobius strip) in the real space. The color indicates the twisting number.}
\end{figure}

\begin{figure}[h]
\begin{center}
\includegraphics*[width=8cm]{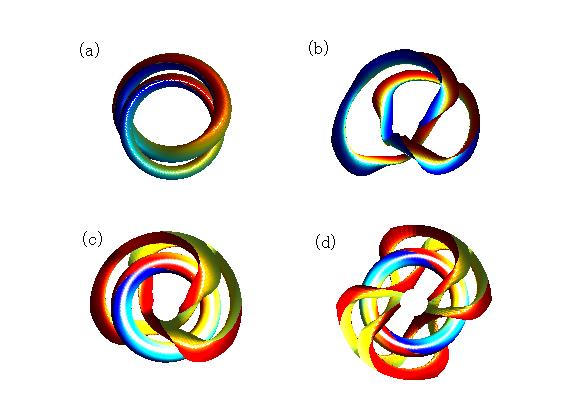}
\end{center}
\caption{(Color online) The isosurfaces correspond to $d_z=0.98$ and $d_x^2+(d_y-d_z)^2/2+0.98^2=1$ (a) or $d_x=0.98$ and $d_y=0.98$ (b) with $n=1$, \textsl{e.g.}, (c) and (d) The isosurfaces correspond to $d_z=0.98$, $d_x=0.98$, $dy=0.98$ for $n=1$ and $n=2$, respectively.}
\end{figure}

In knot theory twisted vortex ring defines an topological unknot, the simplest knot-like structure\cite{Babaev,Faddeev}. The knot can be characterized by the Hopf charge which is defined by\cite{Yuki2}
\begin{equation}
Q_\textrm{H}=\frac{1}{4\pi^2}\int d^3\textbf{r}\varepsilon_{ijk}\mathcal{F}_{ij}\mathcal{A}_k,
\end{equation}
where $\mathcal{F}_{ij}=\partial_i\mathcal{A}_j-\partial_j\mathcal{A}_i={\bf d}\cdot(\partial_i{\bf d}\times\partial_j{\bf d})$ is the strength of the gauge field and the vector $\mathcal{\bf A}({\bf r})$ defines a connection in the order parameter space. As we know, for integer knot, the hopf charge is intepreted as the linking number of preimages\cite{Yuki2,JH}. It is directly calculated that the state (\ref{solution}) has a fractional topological charge of $Q=n/2$, where the factor $1/2$ comes from the semi-sphere instead of the full-sphere. Hence we call it the half-knot so as to differentiate it from the usual integral number of Hopf charge.

The half-knot can also be verified visually. We note that the equilatitude surface and equilongitude surface of the semi-sphere $S^2_{1/2}$ is no longer equivalent as in the integer knot in which the manifold is $S^2$\cite{Yuki2}. Figure 5 schematically shows the two distinct mappings for the twisted tube (isosurface $d_z=0.98$) and the twisted Mobius strip (isosurface $d_x=-0.98$). The color on the tube represents for the value of $\arctan(d_y/d_x)$ while on the Mobius strip the value of $\arctan(d_y/d_z)$, which help to recognize the twisting feature of the tube or the strip. We mention that the color on the semi-sphere is just schematic and has no direct relevance to the color on the tube or the strip.

The topological structures are shown in Fig.6. Fig.(6)(a,b) and (c,d) display the isosurfaces of the tubes or the strips that correspond to two and three small circles or strings on the $S^2_{1/2}$, respectively. In Fig.6(a) the isosurfaces correspond to $d_z=0.98$ and $d_x^2+(d_y-d_z)^2/2+0.98^2=1$. The two twisted tubes are linked exact once, which is the same as the integer knot. In Fig.6(b) the linked strips correspond to the isosurfaces of $d_x=0.98$ and $d_y=0.98$, respectively. Fig.6(c,d) display three isosurfaces from three circles or strings on the $S^2_{1/2}$ by $d_z=0.98$, $d_x=0.98$, and $dy=0.98$, respectively. Two of the tubes or strips are linked once for $n=1$ (Fig.6(a-c)) and twice for $n=2$ (Fig.6(d)).

\section{Summary}
In summary, we have constructed an exact solution to the coupled nonlinear GPEs in the 3D space. We showed that the spinor condensates can accommodate a solitonic knot with a half-integer topological charge. A knot may be created in the spinor condensate by using a quadrupolar magnetic field and observed in a Stern-Gerlach experiment\cite{Ja,Yuki2}. This technique can be extended to create exotic spin textures. Our knot state can be stabilized by an external trap of the shape $V(x)=V_0(\tanh^2(kx)+\tanh^2(ky))$, which coincides with the total density profile. Finally, we mention that our analytical method can be applied to other coupled nonlinear equations. Some relevant works will be published elsewhere.

This work is supported by the funds from the Ministry of Science and Technology of China under Grant No. 2012CB821403.


\begin{thebibliography}{99}
\bibitem{Volovik} G.E. Volovik and V.P. Mineev, Sov. Phys. JETP 45(6), 1186-1196 (1977); D. J. Thouless, Topological Quantum Numbers in Nonrelativistic Physics (World Scientific, Singapore, 1988).
\bibitem{Faddeev} L. Faddeev and A. J. Niemi, Nature 387, 58¨C61 (1997).
\bibitem{Battye} R. A. Battye and P. M. Sutcliffe, Phys. Rev. Lett. 81, 4798 (1998).
\bibitem{Irvine} W. T. M. Irvine and D. Bouwmeester, Nat. Phys. 4, 716-720 (2008).
\bibitem{Dennis} M. R. Dennis, R. P. King, B. Jack, K. O¡¯Holleran, and M. J. Padgett, Nat. Phys. 6, 118-121 (2010).
\bibitem{Kivshar} A. S. Desyatnikov, D. Buccoliero, M. R. Dennis and Y. S. Kivshar, Sci. Rep. 2, 771 (2012).
\bibitem{Babaev} E. Babaev, L. D. Faddeev, and A. J. Niemi, Phys. Rev. B 65, 100512 (2002).
\bibitem{Yuki2} Y. Kawaguchi, M. Nitta, and M. Ueda, Phys. Rev. Lett. 100, 180403 (2008).
\bibitem{Vilenkin} A. Vilenkin and E. P. S. Shellard, Cosmic Strings and Other Topological Defects (Cambridge University Press, Cambridge, England, 1994).
\bibitem{Ueda} M. Ueda and Y. Kawaguchi, Phys. Rep. 500, 253 (2012).
\bibitem{Yuki} Y. Kawaguchi, M. Kobayashi, M. Nitta, and M. Ueda, Prog. Theor. Phys. Suppl. 186, 455-462 (2010).
\bibitem{AE} A. E. Leanhardt, Y. Shin, D. Kielpinski, D. E. Pritchard, and W. Ketterle, Phys. Rev. Lett. 90, 140403 (2003).
\bibitem{LE} L. E. Sadler, J. M. Higbie, S. R. Leslie, M. Vengalattore, and D. M. Stamper-Kurn, Nature 443, 312 (2006).
\bibitem{Martikainen} J.-P. Martikainen, A. Collin, and K.-A. Suominen, Phys. Rev. Lett. 88, 090404 (2002).
\bibitem{Pietila} V. Pietil\"{a} and M. M\"{o}tt\"{o}nen, Phys. Rev. Lett. 103, 030401 (2009).
\bibitem{ND} N. D. Mermin and T. L. Ho, Phys. Rev. Lett. 36, 594 (1976).
\bibitem{TM} T. Mizushima, K. Machida, and T. Kita, Phys. Rev. Lett. 89, 030401 (2002).
\bibitem{PW} P. W. Anderson and G. Toulouse, Phys. Rev. Lett. 38, 508 (1977).
\bibitem{MH2} H. M\"{a}kel\"{a}, J. Phys. A: Math. Gen. 39, 7423-7439 (2006).
\bibitem{Anglin} J. Ruostekoski and J. R. Anglin, Phys. Rev. Lett. 86, 3934 (2001).
\bibitem{Savage} C. M. Savage and J. Ruostekoski, Phys. Rev. Lett. 91, 010403 (2003).
\bibitem{Ruostekoski} J. Ruostekoski, Phys. Rev. A 70, 041601(R) (2004).
\bibitem{Yang1} S.J. Yang, Q.S. Wu, S.N. Zhang, and S. Feng, Phys. Rev. A 77, 033621 (2008).
\bibitem{Cong} C. Zhang, W. Guo, S. Feng, and S.J. Yang, arXiv: 1302.5504 (unpublished).
\bibitem{Ja} J.-Y. Choi, W. J. Kwon, and Y.-I. Shin, Phys. Rev. Lett. 108, 035301 (2012).
\bibitem{HZ} H. Zhai, W. Q. Chen, Z. Xu, and L. Chang, Phys. Rev. A 68, 043602 (2003).
\bibitem{A} A. Saxena, R. Dandoloff, Phys. Rev. B 66, 104414 (2002).
\bibitem{LB} L. Brey, H. A. Fertig, R. Cote, and A. H. MacDonald, Phys. Rev. B 54, 16888-16902 (1996).
\bibitem{Ta} T. Morinari, Phys. Rev. B 72, 104502 (2005).
\bibitem{VM} V. M. H. Ruutu, J. Kopu, M. Krusius, U. Parts, B. Placais, E. V. Thuneberg, and W. Xu, Phys. Rev. Lett. 79, 5058-5061 (1997).
\bibitem{R} R. Ishiguro, O. Ishikawa, M. Yamashita, Y. Sasaki, K. Fukuda, M. Kubota, H. Ishimoto, R. E. Packard, T. Takagi, T. Ohmi, and T. Mizusaki, Phys. Rev. Lett. 93, 125301 (2004).
\bibitem{WM1} S.-W. Su, I.-K. Liu, Y.-C. Tsai, W. M. Liu, and S.-C. Gou, Phys. Rev. A 86 023601 (2012).
\bibitem{MM} M. M. Salomaa and G. E. Volovik, Rev. Mod. Phys 59, 533 (1987).
\bibitem{JH} J. Jaykka and J. Hietarinta, Phys. Rev. D 79, 125027 (2009).
\end{thebibliography}
\end{document}